\documentclass[%
 aip,
 amsmath,amssymb,
 reprint,%
]{revtex4-2}

\usepackage{graphicx}
\usepackage{dcolumn}
\usepackage{bm}
\usepackage{xcolor}
\usepackage[utf8]{inputenc}
\usepackage[T1]{fontenc}
\usepackage{mathptmx}
\usepackage{etoolbox}

\makeatletter
\def\@email#1#2{%
 \endgroup
 \patchcmd{\titleblock@produce}
  {\frontmatter@RRAPformat}
  {\frontmatter@RRAPformat{\produce@RRAP{*#1\href{mailto:#2}{#2}}}\frontmatter@RRAPformat}
  {}{}
}%
\makeatother
\begin{document}

\preprint{AIP/123-QED}

\title{Experimental Study of Fabry–Pérot BICs in a Microwave Waveguide}
\author{Zilong~Zhao}
\affiliation{Qingdao Innovation and Development Center of Harbin Engineering University, 266404, Qingdao, China}

\author{Nikolay~Solodovchenko}
\affiliation{School of Physics and Engineering, ITMO University, 197101, St. Petersburg, Russia}
\author{Chao~Sun}
\affiliation{Qingdao Innovation and Development Center of Harbin Engineering University, 266404, Qingdao, China}
\author{Mingzhao~Song}
\affiliation{Qingdao Innovation and Development Center of Harbin Engineering University, 266404, Qingdao, China}
\email{kevinsmz@foxmail.com} 
\author{Ekaterina~Maslova}
\affiliation{School of Physics and Engineering, ITMO University, 197101, St. Petersburg, Russia}
\email{ekaterina.maslova@metalab.ifmo.ru}
\author{Andrey~Bogdanov}
\affiliation{Qingdao Innovation and Development Center of Harbin Engineering University, 266404, Qingdao, China}%
\affiliation{School of Physics and Engineering, ITMO University, 197101, St. Petersburg, Russia}
\email{a.bogdanov@hrbeu.edu.cn}
\date{\today}

\begin{abstract}
We study Fabry–Pérot bound states in the continuum (FP-BIC) in the GHz frequency range, formed by two ceramic discs placed inside a metallic-walled rectangular waveguide, that act as perfect reflectors at the resonant frequency. The energy becomes perfectly trapped between the discs, forming a FP-BIC, when the distance between them matches the Fabry–Pérot quantization condition. We present both theoretical and experimental analyses to investigate how the total and radiative quality factors (Q factors) depend on the inter-disk distance. We gain valuable insights into the Fano features observed in the transmission spectra using the quasi-normal mode technique and temporal coupled mode theory. Notably, we find that as the system approaches the BICs, the Fano asymmetry parameters diverge, resulting in a Lorentzian transmission profile. Experimentally, we measure a radiative Q factor on the order of $10^5$, while the total Q factor, limited by material losses, remains around $10^3$. These results offer new opportunities for the application of BICs in microwave technology, significantly advancing the potential for high-performance devices.
\end{abstract}

\maketitle


{\it Bound states in the continuum} (BIC) represent remarkable nonradiative eigenmodes that, despite their energy residing within the continuum spectrum, remain perfectly localized without coupling to radiative channels. These states exhibit infinite radiative quality factors (Q factors) in ideal, lossless systems and have attracted significant interest due to their ability to enhance light–matter interactions substantially. In photonics and microwave technology\cite{hsu2016bound,koshelev2019nonradiating,azzam2021photonic,sadreev2021interference,koshelev2023bound}, BICs offer unprecedented opportunities for high-performance resonators due to their strong field confinement~\cite{koshelev2020subwavelength,zalogina2023high,mylnikov2020lasing}, tunability~\cite{xu2019dynamic,hu2022spatiotemporal}, and robustness to perturbations~\cite{kuhne2021fabrication,jin2019topologically,maslova2021bound}.

The unique characteristics of BICs have already enabled numerous applications in various domains, such as laser systems~\cite{zhang2022quasi,asada1986gain,hwang2021ultralow,kodigala2017lasing,wang2020generating,cui2018multiple}, optical sensors~\cite{romano2018surface,yin2006goos,zhang2023enhanced,hu2023high}, nonlinear optical applications~\cite{kivshar2017meta,koshelev2019nonlinear,krasikov2018nonlinear,carletti2018giant,bulgakov2019nonlinear}, and filtering devices~\cite{foley2014symmetry,foley2015normal,cui2016normal}. Specifically, their ability to achieve ultrahigh Q factors and localized electromagnetic fields makes them exceptionally useful for precise frequency selection, enhanced nonlinear interactions, ultrasensitive biosensing, and advanced photonic integrated circuits~\cite{yu2019photonic}. Additionally, BICs exhibit intriguing topological properties~\cite{zhen2014topological,kang2022merging,xiao2017topological}, including polarization vortices characterized by quantized topological charges, which make them valuable in studies of the generation of beams with angular momentum and vortex lasing.

Among the known mechanisms for BIC formation, widely discussed in the literature~\cite{ndangali2010electromagnetic,ndangali2013resonant,bulgakov2018propagating,hemmati2019resonant,shuai2013double,shuai2013coupled,maslova2021bound}, Friedrich–Wintgen BICs stand out as one of the most general types. These BICs arise from the destructive interference of two leaky resonances, whose radiative losses cancel exactly under specific conditions. Crucially, the formation of Friedrich–Wintgen BICs requires fine-tuning system parameters to ensure complete destructive interference, typically leading to an avoided resonance crossing where one mode becomes infinitely long-lived while its counterpart radiates strongly.

\begin{figure}
\centering
\includegraphics[width = 1\linewidth]{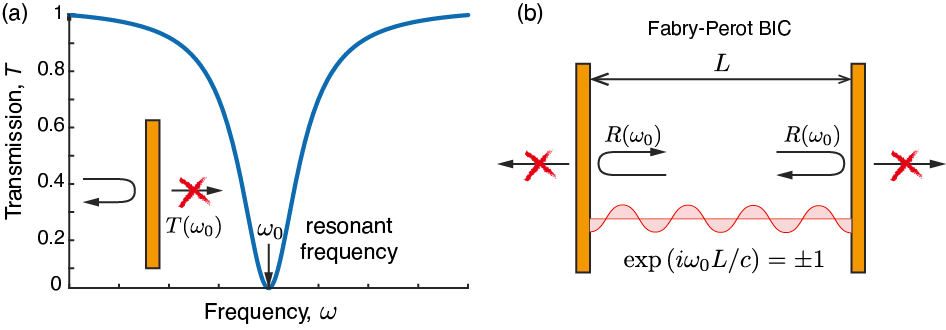} 
\caption{(a) Transmission spectrum of a single resonant mirror exhibiting a resonance at frequency $\omega_0$. The inset illustrates the schematic geometry of the mirror.  
(b) Schematic illustration of a Fabry-Pérot bound state in the continuum formed between two identical resonant mirrors.}
\label{fig:Intro}
\end{figure}

A specific and illustrative subclass of Friedrich–Wintgen BICs is the Fabry–Pérot (FP) BIC~\cite{plotnik2011experimental,koshelev2018asymmetric}, arising when radiation is trapped between two resonant mirrors as shown in Fig.~\ref{fig:Intro}. Physically, each resonant mirror acts as a perfect reflector at its resonant frequency $\omega_0$ [see Fig.~\ref{fig:Intro}(a)], and by placing two such mirrors face-to-face, one can confine electromagnetic waves between them. This confinement occurs precisely when the distance between the mirrors $L$ satisfies the Fabry–Pérot quantization condition $\exp(i\omega_0L/c)=\pm1$, ensuring a round-trip phase shift equal to an integer multiple of 2$\pi$ [Fig.~\ref{fig:Intro}(b)]. The formation of FP BIC is shown schematically in Fig.~\ref{fig:Intro}.

\begin{figure}
\centering
\includegraphics[width = 1\linewidth]{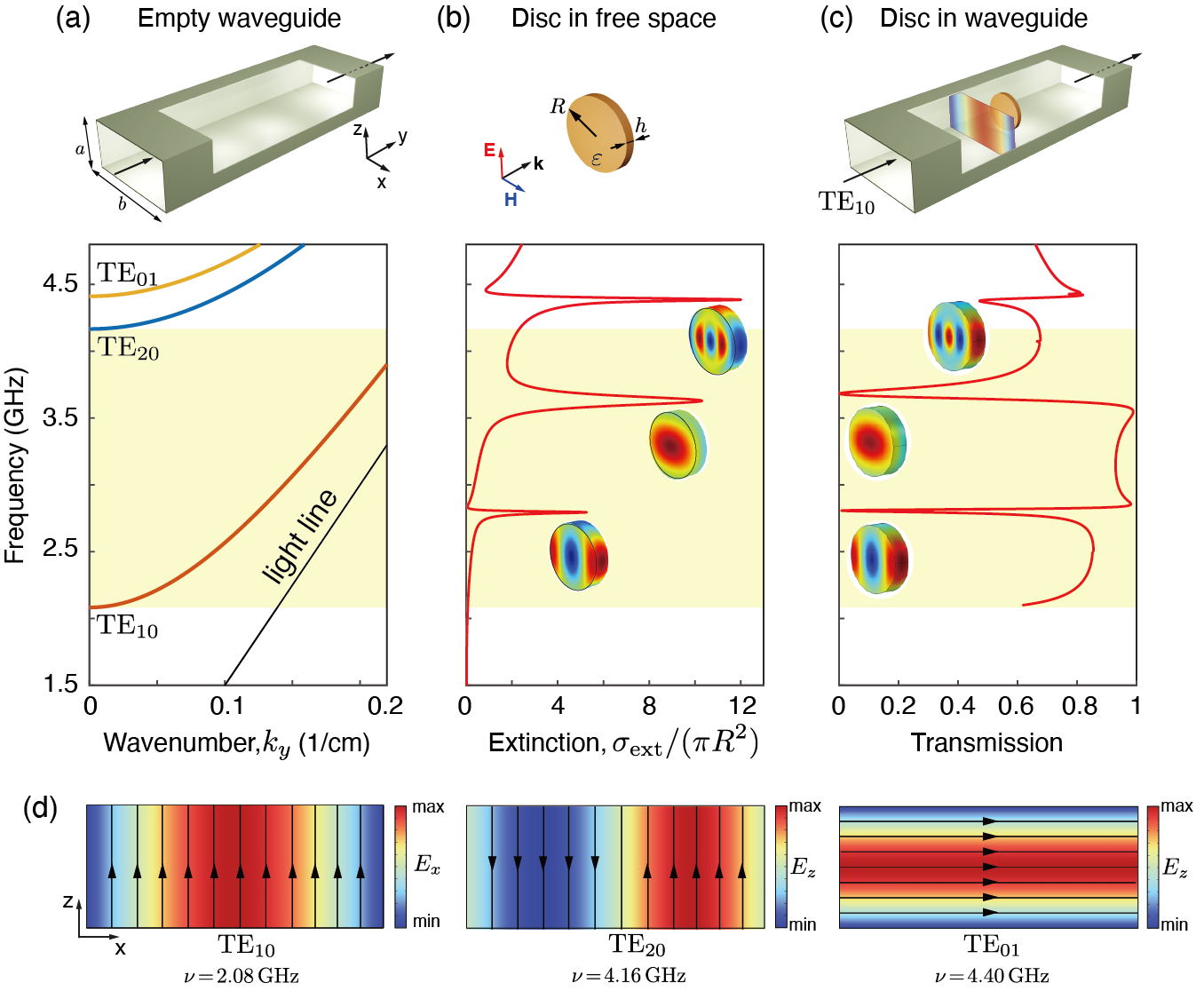} 
\caption{(a) Geometry of the rectangular waveguide with cross-sectional dimensions $a = 34$~mm and $b = 72$~mm (upper panel). The lower panel shows the dispersion relations of the three lowest-order waveguide modes. The shaded yellow region indicates the frequency range corresponding to single-mode (TE$_{10}$) operation.  
(b) Normalized extinction cross-section of a ceramic disc in free space (lower panel). The insets display the $E_z$-field distribution on the disc surface for the resonant eigenmodes corresponding to the peaks in the extinction spectra. The disc parameters are shown in the upper panel: radius $R = 10$~mm, thickness $h = 5$~mm, permittivity $\varepsilon = 80$, and loss tangent $\tan\delta = 10^{-3}$.   (c) Transmission spectrum of the TE$_{10}$ mode through the waveguide loaded with the ceramic disc (lower panel). The insets illustrate the $E_z$-field profile of the disc eigenmode inside the waveguide. The upper panel shows the corresponding system geometry.}
\label{fig:mode}
\end{figure}

The first study explicitly considering FP-BICs in photonics was conducted by Marinica and Borisov, who explored FP-BIC formation between two dielectric gratings~\cite{marinica2008bound}. Almost at the same time, Bulgakov and Sadreev theoretically studied FP-BIC in a photonic crystal waveguide with two defects interacting with a guided continuum~\cite{bulgakov2008bound}. Today, FP-BICs are quite well studied theoretically~\cite{alagappan2024fabry}. Gao et al.~\cite{gao2016formation} elucidated how guided resonances and BICs arise from FP-type standing waves in photonic crystal slabs, identifying conditions for both symmetry-protected and accidental BICs. Nabol et al.~\cite{nabol2022fabry} presented analytical conditions for FP-BIC formation in anisotropic photonic crystals, explicitly linking phase-matching and interference processes. Mai and Lu~\cite{mai2025relationship} provided insights into the relationship between total reflection and FP-BICs, emphasizing the necessity of specific mirror symmetries for achieving perfect confinement. Additionally, Shubin et al.~\cite{shubin2023twin} introduced the concept of twin FP-BICs arising when mirrors themselves support BICs, allowing for two degenerate nonradiative modes almost independent of cavity length. Recently, Ni et al.~\cite{ni2024three}demonstrated tunable FP-BICs using bilayer photonic crystals, achieving dynamic modulation of BICs and exploiting interlayer coupling to control optical singularities such as spatiotemporal phase singularities and orbital angular momentum. FP-BICs are currently under active experimental research. Thus, Rezzouk et al.~\cite{rezzouk2024fabry} demonstrated FP and Friedrich–Wintgen BICs experimentally in a triple-stub microwave resonator, highlighting how symmetric stub arrangements can yield FP-BICs through precise phase conditions, while asymmetric configurations enable Friedrich–Wintgen interference-driven BIC formation. Wu et al.~\cite{song2024high} observed FP-BICs in the near-infrared regime using bilayer dielectric gratings, where tuning inter-grating spacing enabled control over BIC positions in momentum space and enhanced nonlinear optical interactions. Rao et al.~\cite{rao2011study} fabricated visible-range FP cavities employing dielectric metasurfaces suspended over Bragg mirrors, confirming the theoretical relationship that the Q factor diverges inversely with the square of the detuning from the BIC conditions.

FP-BICs appear in effectively one-dimensional systems, which are challenging to realize experimentally due to practical constraints. For example, using metasurfaces or gratings as perfect mirrors requires their lateral sizes to be effectively infinite. However, a waveguide with metallic walls offers an excellent example of the system described by a one-dimensional effective Hamiltonian~\cite{sadreev2021interference}. Lepetit and Kanté have shown that microwave waveguides provide feasible experimental platforms for observing and analyzing BICs~\cite{lepetit2014controlling}.

In this paper, we experimentally study FP-BICs within the GHz frequency range, realized by placing two ceramic discs inside a metallic-walled rectangular waveguide. Acting as nearly perfect reflectors, these discs trap radiation at discrete inter-disc distances defined by the Fabry–Pérot quantization conditions. Combining theoretical analysis and experimental measurements, we investigate the dependence of the total and radiative Q factors on the interdisc spacing. We analyze the formation of pronounced Fano resonances in transmission spectra through the quasi-normal mode technique and temporal coupled-mode theory. Intriguingly, we find that as the configuration approaches the FP-BIC conditions, the Fano asymmetry parameters diverge, leading to nearly perfect Lorentzian transmission profiles. Our results pave the way for new microwave technologies harnessing the unique properties of BICs, thereby significantly advancing the potential for developing high-performance devices.

\begin{figure}
\centering
\includegraphics[width = 1.0\linewidth]{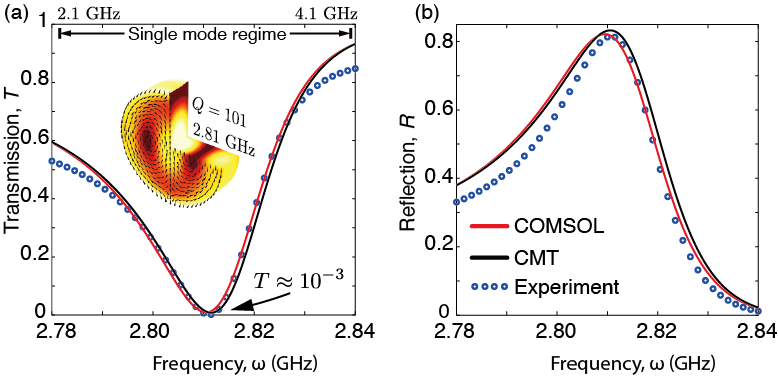} 
\caption{(a) Transmission and (b) reflection spectra of the ceramic disc placed inside a waveguide. The solid red lines represent numerical results obtained using COMSOL Multiphysics, the black lines correspond to temporal coupled-mode theory (CMT) simulations, and the blue circles indicate experimental measurements. Inset in panel (a) show the electric field distributions at the resonant frequencies and the $Q$-factor of the resonance mode.}
\label{fig:OneDisk}
\end{figure}

\begin{figure*}
\centering
\includegraphics[width = 1\linewidth]{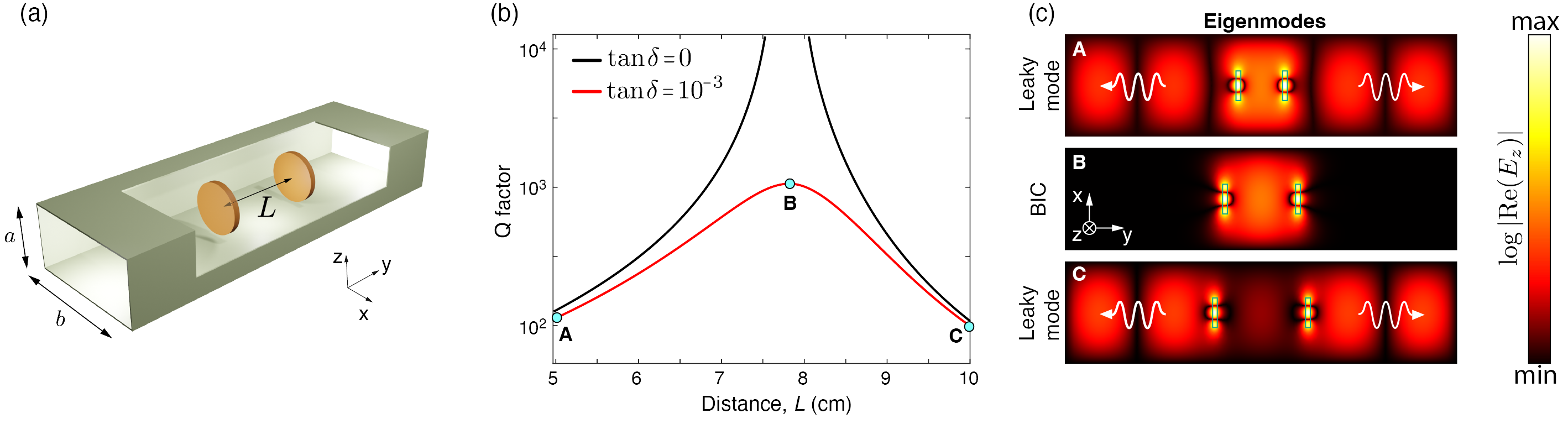} 
\caption{(a) Schematic view of two disks in a waveguide. (b) The Q-factor depends on the distance between two disks with or without losses. (c) Electric field at different distances}
\label{fig:TwoDisks}
\end{figure*}

We consider a rectangular microwave waveguide (WR-284) with a cross-sectional height of $a = 34~\mathrm{mm}$ and width of $b = 72~\mathrm{mm}$. The dispersion relations of the three lowest-order modes (TE$_{10}$, TE$_{20}$, and TE$_{01}$) are presented in Fig.~\ref{fig:mode}(a). The electric field distribution for these modes is shown in Fig.~\ref{fig:mode}(d). The waveguide supports single-mode operation between the cutoff frequencies of the TE$_{10}$ and TE$_{20}$ modes, which are 2.08~GHz and 4.17~GHz, respectively, while the recommended operating frequency range is 2.60–3.95~GHz. 

Figure~\ref{fig:mode}(b) presents the normalized extinction cross-section spectrum of the ceramic Mie-resonant disc in free space, illuminated by a plane wave propagating along the disc’s rotational axis, as illustrated in the upper panel of the figure. The insets show the electric field profiles of the Mie-modes. The extinction spectrum and mode profiles are calculated in COMSOL Multiphysics. The disc has the following parameters: dielectric permittivity $\varepsilon = 80$, loss tangent $\tan\delta = 10^{-3}$, radius $R = 10$~mm, and thickness $h = 5$~mm. 

Figure~\ref{fig:mode}(c) shows the numerical transmission spectrum of the TE$_{10}$ mode through a rectangular waveguide loaded with the ceramic disc positioned at its center. The simulation is performed using COMSOL Multiphysics. The pronounced drop in transmission at resonance frequencies indicates that the disc inside the waveguide serves as a highly reflective resonant mirror. The insets in Fig.~\ref{fig:mode}(c) display the field distributions of the corresponding eigenmodes of the discs within the waveguide, which remain qualitatively similar to those in the free-space, suggesting that the waveguide introduces only weak perturbations to the mode structure. However, the waveguide does play a critical role in symmetry reduction. Notably, there are two shallow dips in Fig.~\ref{fig:mode}(c) near 4.1 and 4.4~GHz. They correspond to free-space Mie modes with C$_3$ rotational symmetry, and their excitation is prohibited in the configuration shown in Fig.~\ref{fig:mode}(b) due to selection rules~\cite{gladyshev2024fast,gladyshev2020symmetry}. However, the waveguide lowers the system symmetry from D$_{\infty h}$ to D$_{4h}$ relaxing the selection rules.

Figures~\ref{fig:OneDisk} presents the transmission and reflection spectra of a ceramic disc inside a rectangular waveguide, analyzed both numerically using COMSOL Multiphysics (red solid lines), analytically using temporal CMT (black solid lines), and experimentally (blue open circles). Panel~(a) shows the transmission spectrum $T(\omega)$ near the resonance frequency 2.81~GHz, where the disc acts as a nearly perfect mirror, with transmittance dropping to $T \approx 10^{-3}$ (the measured value). The inset in panel~(a) displays the electric field distribution of the resonant mode and its quality factor $Q = 101$. The CMT was used to fit the experimental data and extract the experimental value of the loss tangent, which was found to be $\tan\delta = 10^{-3}$. {This value is consistent with the material specifications C80R-A provided by the manufacturer Cai Qin Technology.} This extracted value was further cross-validated by full-wave COMSOL simulations, confirming the consistency of the model and the extracted value of the loss tangent. Panel~(b) shows the corresponding reflection spectrum $R(\omega)$, which exhibits a peak near the same resonant frequency. The experimental reflection spectrum  agrees well with the CMT model and numerical simulation. The reflection reaches a maximum of approximately 0.8 due to resonant absorption in the disc. However, this does not impose a fundamental limit on the Q-factor of FP-BICs, as the amount of energy stored can be made arbitrarily large by increasing the distance between the resonant mirrors~\cite{semushev2025robustness}.

Figure~\ref{fig:TwoDisks} illustrates the emergence of FP-BICs formed by two ceramic discs placed inside a rectangular waveguide. Panel~(a) presents a schematic of the waveguide with two identical dielectric discs separated by a distance $L$, which serves as a tuning parameter. Panel~(b) shows how the $Q$ factor of the resonant mode depends on the inter-disc distance, both in the absence of losses (black curve, $\tan\delta = 0$) and in the presence of material absorption (red curve, $\tan\delta = 10^{-3}$). The Q-factor reaches a pronounced maximum at $L = 7.8$~cm, indicating the emergence of a FP-BIC where the radiative losses are suppressed due to the extremely low transmission through the discs. This is evident from the divergent peak in the $Q$ factor in the lossless case, which is reduced to $Q\approx10^3$ in the presence of material absorption. Points A and C in the plot correspond to detuned distances where the Fabry–Pérot quantization condition is not met and the structure supports only leaky modes with significantly lower Q-factors. Panel~(c) visualizes the real part of the electric field $E_z$ in logarithmic scale for these three cases. At point A, the field leaks out of the structure, forming a low-Q leaky mode. At point B, the field is fully confined between the discs, confirming the presence of a BIC. At point C, the resonance condition is again violated, resulting in a low-Q leaky mode. The field maps clearly show the difference in confinement behavior, with strong localization only at the BIC condition.

\begin{figure*}
\centering
\includegraphics[width = 1\linewidth]{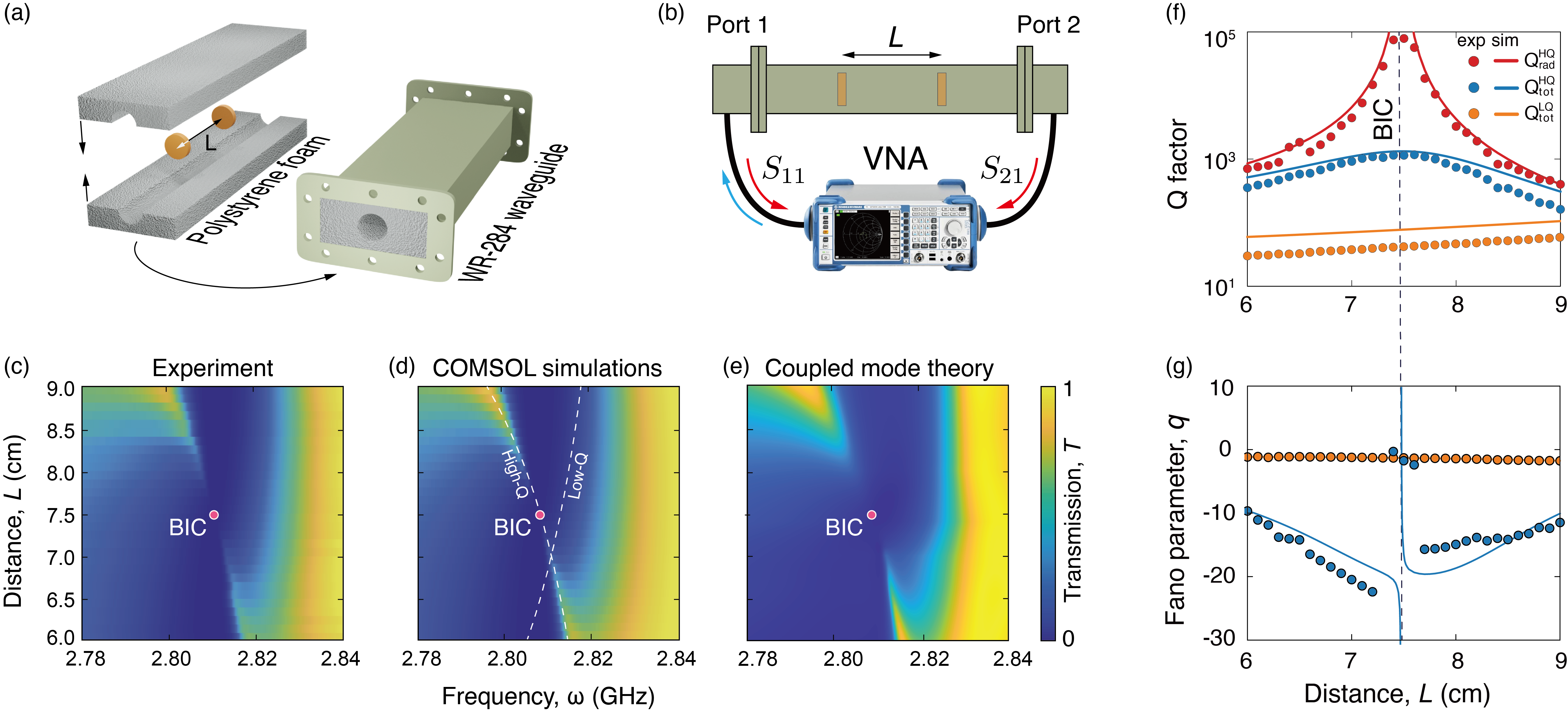} 
\caption{(a) Schematic diagram of the experiment equipment. (b) Schematic diagram of the experiment setup connection.(c) Transmission map of the experiment results. (d) Transmission map of the simulation results. (e) Transmission map of the coupled mode theory results.  (f) Simulation calculation and experiment fitting of QNMs' Q-factor at different distances. (g) Simulation calculation and experiment fitting of QNMs'Fano parameter at different distances.}
\label{fig:Comparison}
\end{figure*}

For the experimental study of the FP-BIC in the microwave waveguide, we take two identical ceramic discs symmetrically positioned inside a WR-284 rectangular waveguide using polystyrene foam holders, as shown in Fig.~\ref{fig:Comparison}(a). The relative permittivity of the polystyrene foam used in the setup is estimated to be $\varepsilon = 1.04$.  The distance between the discs $L$ is adjustable and plays a key role in tuning the BIC conditions. The distance $L$ is measured before setting the discs inside the waveguide, ensuring accurate alignment. A schematic of the experimental setup is shown in Fig.~\ref{fig:Comparison}(b), where a vector network analyzer (VNA) is connected to the two ports of the waveguide to measure the complex reflection ($S_{11}$) and transmission ($S_{21}$) coefficients.

The experimental results are presented in Fig.~\ref{fig:Comparison}(c), where the transmission spectrum is shown as a function of frequency and inter-disc distance $L$, which was varied from 6~cm to 9~cm in steps of 0.1~mm. A sharp transmission minimum is observed near $L\approx7.8$~cm, indicating the formation of a FP-BIC. This value of $L$ is slightly shifted compared to that in Fig.~\ref{fig:TwoDisks}(b), which can be attributed to the perturbative influence of the polystyrene foam holders. The FP-BIC is marked in the transmission map by a red circle marker. This observation is supported by full-wave simulations using COMSOL Multiphysics [Fig.~\ref{fig:Comparison}(d)]. The two white dashed lines indicate the high-Q and low-Q resonant modes. The FP-BIC generally emerges in a pair with a low-Q companion mode, associated with a Fabry–Pérot resonance of opposite parity in terms of number of half-wavelengths confined between the mirrors~\cite{semushev2025robustness}. Analytical modeling using temporal CMT [Fig.~\ref{fig:Comparison}(e)] agrees with the numerical and experimental results. The CMT model is used to fit the experimentally measured spectra of S-parameters.The near-field coupling between the discs $\kappa$ and the elements of the direct scattering matrix $\hat{C}$ were treated as fitting parameters.

To describe the resonance properties of the system in detail, we used the formalism of quasi-normal mode theory (QNM), which is a solution of Maxwell's equations without sources or eigenmodes. To solve the eigenvalues of the system, normalize the modes, and calculate the extinction cross-section, we used "MAN" (Model Analysis of Nanostructures)~\cite{lalanne2018light}, where the waveguide mode of a rectangular waveguide was used as the incident wave. The extinction cross-section is a linear sum of QNMs:

\begin{equation}
    \sigma^\text{ext}= \sigma^\text{non} + \sum_m \sigma^\text{ext}_m,
\end{equation}
where $\sigma^\text{non}$ is the non-resonant part and $\sigma^\text{ext}_m$ is the response of one QNM, which can be represented as follows~\cite{bochkarev2024quasinormal}:

\begin{equation}
    \sigma^\text{ext}_m = \sigma_m \left[\frac{q_m^2 - 1 + 2q_m \Omega_m}{(\Omega_m^2+1)(q_m^2+1)}   \right],
\end{equation}
where $ \sigma_m $ and $q_m$ are the intensity and Fano parameter of the QNM mode, respectively; $\Omega_m = (\omega-\omega_m)/\Gamma_m$ is the dimensionless frequency; $\Gamma_m$ and $\omega_m$ are the resonance width and frequency, respectively. In case of small losses in the resonator $\sigma^\text{non} \approx 0$.

According to Ref.~\onlinecite{bochkarev2024quasinormal}, the intensity $\sigma_m$ and Fano asymmetry parameter $q_m$ can be found using the overlap integrals between the QNM and the incident field, which in our case is an analytically defined waveguide mode. The extinction cross-section in the experiment can be obtained using the S-parameters with the help of the optical theorem:

\begin{equation}
    \sigma_\text{ext}=2\mathrm{Re}(1-S_{12}^\prime),
\end{equation}
where $S_{12}^\prime = S_{12} e^{i \beta_{10} d}$ is the scattering matrix element obtained in the experiment with an additional phase factor depending on the waveguide length; $\beta_{10}$ is the propagation constant of the TE$_{10}$ waveguide mode; $d$ is the length of the waveguide.

Thus, the experimental spectrum of the extinction cross-section can be fitted using the parameters (resonance frequency $\omega_m$, damping $\Gamma_m$, intensity $\sigma_m$, and Fano parameter $q_m$) of each QNM as an initial approximation of the experimental parameters. In the frequency region under study, we used the parameters of two QNMs, and the others were approximated by a linear function. Therefore, we can fit the experiment data and get the actual results of the quality factors and Fano parameters [see Figures~\ref{fig:Comparison}(f) and~\ref{fig:Comparison}(g)].

By fitting the experimental extinction cross-section spectra, we obtain the experimental resonance frequencies $\omega_m$ and the total quality factors $Q_\text{tot}$ of each resonance. Using the experimental S-parameters and the obtained quality factor ($Q_\text{tot}$), we can obtain the radiative part of the quality factor

\begin{equation}
    Q_\text{rad}=Q_\text{tot}\left(1- \left| \frac{2 S_{11}}{S_{12}} \right|  \right),
\end{equation}
which diverges at the BIC. It is important to note that the Fano asymmetry parameter $q_m$ also tends to $\pm~\infty$ when the distance between cylinders is close to 7.5~cm, which was previously also observed in single resonators~\cite{koshelev2023bound}. It appears as a Lorentz line in the extinction cross-section spectrum. This behavior of the Fano parameter $q_m$ is because the resonance loses its connection with the continuum.

In summary, we have experimentally and theoretically investigated the formation of Fabry–Pérot bound states in the continuum (FP-BICs) in a rectangular metallic waveguide loaded with two ceramic Mie-resonant discs. By varying the inter-disc distance, we observed the emergence of FP-BICs when the Fabry–Pérot quantization condition is satisfied, leading to the complete suppression of radiative losses. We demonstrated that the radiative $Q$ factor can exceed $10^5$, while the total $Q$ factor is limited by material absorption and reaches values around $10^3$. Our analysis, based on temporal coupled-mode theory and quasi-normal mode expansion, reveals that the Fano asymmetry parameter diverges near the BIC point, indicating a transition from Fano to Lorentzian line shapes in the transmission spectra. Theoretical predictions are in excellent agreement with full-wave COMSOL simulations and experimental measurements. These findings provide a robust experimental platform for the realization of high-$Q$ microwave resonators and open new pathways for implementing BIC-enabled functionalities in sensing, filtering, and signal processing applications at GHz frequencies.

\bibliography{ref}

\end{document}